\documentclass[a4paper,12pt]{article}

\usepackage[T1]{fontenc}
\usepackage{ae,aecompl}
\usepackage{color}
\usepackage{geometry}

\usepackage{graphicx}	% Including figure files

\usepackage{subcaption}
\usepackage{float}
\usepackage[fleqn]{amsmath}	% Advanced maths commands
\usepackage{amssymb}	% Extra maths symbols
\usepackage{gensymb}
\usepackage{hyperref}
\usepackage[super]{natbib} %trying to make it like nature
\usepackage[nolist]{acronym}
\acrodef{GW}{gravitational wave}
\acrodef{NS}{neutron star}
\acrodef{BH}{black hole}
\acrodef{BBH}{binary black hole}
\acrodef{aLIGO}{Advanced Laser Interferometer Gravitational-wave observatory}
\acrodef{AdV}{Advanced Virgo}
\acrodef{PN}{post-Newtonian}
\acrodef{BNS}{binary neutron star}
\acrodef{MS}{main-sequence}
\acrodef{HG}{Hertzsprung-Gap}
\acrodef{CHeB}{core helium-burning}
\acrodef{HeMS}{helium star}
\acrodef{HeHG}{post-HeMS star} %\acrodef{HeHG}{helium Hertzsprung-gap}
\acrodef{HeGB}{helium giant}
\acrodef{WR}{Wolf-Rayet}
\acrodef{PE}{parameter estimation}
\acrodef{SNR}{signal-to-noise ratio}
\acrodef{IMF}{initial mass function}
\acrodef{CE}{common-envelope}
\acrodef{MCMC}{Markov-chain Monte Carlo}
\acrodef{LBV}{Luminous Blue Variable}
\acrodef{SMC}{Small Magellanic Cloud}
\acrodef{COMPAS}{Compact Object Mergers: Population Astrophysics and Statistics}
\acrodef{ZAMS}{zero-age main-sequence}

\setcitestyle{comma}

\newcommand{\Msol}[0]{\ensuremath{M_\odot}}
\newcommand{\solar}[0]{\ensuremath{Z_\odot}} % 0.02
\newcommand{\fivepercentsolar}[0]{\ensuremath{5\% Z_\odot}}   % 0.001
\newcommand{\tenpercentsolar}[0]{\ensuremath{10\% Z_\odot}}  % 0.002

\newcommand{\response}[1]{\textcolor{black}{#1}}

\begin{document}

\title{Formation of the first three gravitational-wave observations through isolated binary evolution}

\author{\parbox{\textwidth}{
Simon Stevenson,$^{1}$\thanks{E-mail: simon.stevenson@ligo.org} 
Alejandro Vigna-G\'omez,$^{1}$
Ilya Mandel,$^{1}$\\
Jim W.~Barrett,$^{1}$
\response{Coenraad J.~Neijssel},$^{1}$
David Perkins,$^{1}$
Selma E.~de Mink$^{2}$}
\vspace{0.2cm}\\
% List of institutions
\small\parbox{\textwidth}{$^{1}$ School of Physics and Astronomy, University of Birmingham, Edgbaston, Birmingham B15 2TT, United Kingdom}\\
\small\parbox{\textwidth}{$^{2}$ Anton Pannekoek Institute for Astronomy, University of Amsterdam, 1090 GE Amsterdam, The Netherlands}
}

% Don't change these lines
%\begin{document}
\label{firstpage}
\maketitle

\begin{abstract}
\large During its first 4 months of taking data, Advanced LIGO has detected gravitational waves from two binary black hole mergers, GW150914 and GW151226, along with the statistically less significant binary black hole merger candidate LVT151012. We use our rapid binary population synthesis code COMPAS to show that all three events can be explained by a single evolutionary channel -- classical isolated binary evolution via mass transfer including a common envelope phase.  \response{We show all three events could have formed in low-metallicity environments (Z = 0.001) from progenitor binaries with typical total masses $\gtrsim 160 M_\odot$, $\gtrsim 60 M_\odot$ and $\gtrsim 90 M_\odot$, for GW150914, GW151226, and LVT151012, respectively.}
\end{abstract}

%%%%%%%%%%%%%%%%%%%%%%%%%%%%%%%%%%%%%%%%%%%%%%%%%%

%%%%%%%%%%%%%%%%% BODY OF PAPER %%%%%%%%%%%%%%%%%%

\section*{Introduction}
\label{sec:Introduction}

The \ac{aLIGO}\cite{AdvLIGO} has confidently observed \acp{GW} from two \ac{BBH} mergers, GW150914 \citep{GW150914} and GW151226 \citep{GW151226}.  The \ac{BBH} merger candidate LVT151012 is less statistically significant, but has a $> 86\%$ probability of being astrophysical in origin \citep{GW150914:CBC,BBH:O1}. 

GW150914 was a heavy \ac{BBH} merger, with a well-measured total mass $M = m_1 + m_2 = 65.3 \pm^{4.1}_{3.4}\, M_\odot$ \citep{GW150914:PE,BBH:O1}, where $m_{1,2}$ are the component masses.  Several formation scenarios could produce such heavy \acp{BBH}.  These include: the classical isolated binary evolution channel we discuss in this paper \citep{Belczynski:2016, EldridgeStanway:2016, Lipunov:2016}, including formation from population III stars \citep{Inayoshi:2016}; formation through chemically homogeneous evolution in very close tidally locked binaries \citep{MandeldeMink:2016,Marchant:2016,deMinkMandel:2016}; dynamical formation in globular clusters \citep{Rodriguez:2016,OLeary:2016,Askar:2016}, young stellar clusters \citep{Mapelli:2016}, or galactic nuclei \citep{Bartos:2016dgn,Stone:2016}; or even mergers in a population of primordial binaries \citep{Bird:2016,Sasaki:2016}.  One common feature of all GW150914 formation channels with stellar-origin black holes is the requirement that the stars are formed in sub-solar metallicity environments in order to avoid rapid wind-driven mass loss which would bring the remnant masses below $30 M_\odot$\citep{Belczynski:2010,Spera:2015}; see Results and Abbott \textit{et al.} \cite{GW150914:astro,BBH:O1} for further discussion.

We are developing a platform for the statistical analysis of observations of massive binary evolution, \ac{COMPAS}.  \ac{COMPAS} is designed to address the key problem of \ac{GW} astrophysics: how to go from a population of observed sources to understanding uncertainties about binary evolution.  In addition to a rapid population synthesis code developed with model-assumption flexibility in mind, \ac{COMPAS} also includes tools to interpolate model predictions under different astrophysical model assumptions, astrostatistics tools for population reconstruction and inference in the presence of selection effects and measurement certainty, and clustering tools for model-independent exploration.  

Here, we attempt to answer the following question: can all three LIGO-observed \acp{BBH} have formed through a single evolutionary channel?   We use the binary population synthesis element of \ac{COMPAS} to explore the formation of the observed systems through the classical isolated binary evolution channel \citep{PostnovYungelson:2014} via a \ac{CE} phase \citep{Ivanova:2013}.  \response{We show that GW151226 and LVT151012 could have formed through this channel in an environment at $Z=\tenpercentsolar$ (with $\solar \equiv 0.02$) from massive progenitor binaries with a total \ac{ZAMS} mass $\gtrsim 65 M_\odot$ and $\gtrsim 95 M_\odot$, respectively.}

\response{These \acp{BBH} could also originate from lower-mass progenitors with total masses $\gtrsim 60 M_\odot$ and $\gtrsim 90 M_\odot$, respectively, at metallicity $Z= \fivepercentsolar$, where the same channel could have formed GW150914 from binaries with a total \ac{ZAMS} mass $\gtrsim 160 M_\odot$.  At low metallicity, this channel can produce merging \acp{BBH} with significantly unequal mass ratios: more than $50\%$ of \acp{BBH} have a mass ratio more extreme than 2 to 1 at $Z= \tenpercentsolar$.}

\section*{Results}
\subsection*{Forming GW151226 and LVT151012}
\label{sec:results}

For relatively low-mass \ac{GW} events the \ac{GW} signal in the \ac{aLIGO} sensitive frequency band is inspiral-dominated and the chirp mass $\mathcal{M} = M q^{3/5} (1+q)^{-6/5}$ is the most accurately measured mass parameter, while the mass ratio $q = m_2 / m_1$ cannot be measured as accurately (see figure 4 of \citet{BBH:O1}). The 90\% credible intervals on these for GW151226 and LVT151012 are $8.6 \leq \mathcal{M}/M_\odot \leq 9.2$, $q \geq 0.28$; and $14.0 \leq \mathcal{M}/M_\odot \leq 16.5$, $q \geq 0.24$, respectively \citep{BBH:O1}. For more massive events, the ringdown phase of the \ac{GW} waveform makes a significant contribution and the most accurately measured mass parameter is the total mass $M$.  For GW150914, $M = 65.3 \pm^{4.1}_{3.4}\, M_\odot$ \citep{GW150914:PE,BBH:O1}, with mass ratio $q \geq 0.65$.

\begin{figure*}
	\centering
	\includegraphics[width=\textwidth]{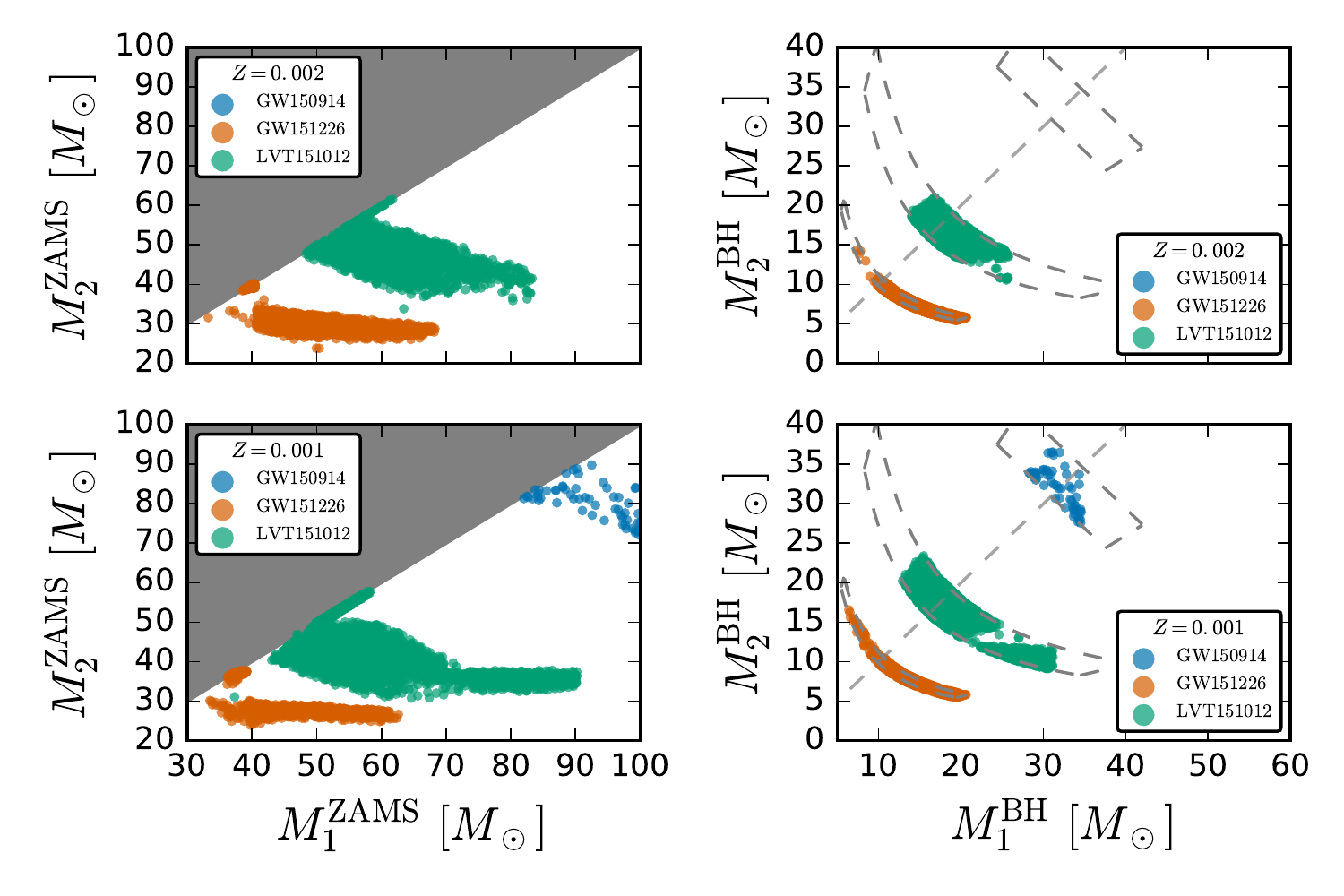}
	\caption[]{\textbf{Masses of binary black holes observed by aLIGO and their progenitors.} 

Each point in the plots represents one system in our simulations. 

(a) \ac{ZAMS} masses $M_1^\mathrm{ZAMS}$ and $M_2^\mathrm{ZAMS}$ for GW150914 (blue - no events), GW151226 (orange) and LVT151012 (green) progenitors at \response{$Z = 10\% \solar = 0.002$}. We define $M_1^\mathrm{ZAMS} > M_2^\mathrm{ZAMS}$ and so shade the non-allowed region gray. 
	
(b) Final black hole masses $M_1^\mathrm{BH}$ and $M_2^\mathrm{BH}$ for merging \acp{BBH} consistent with GW150914, GW151226 and LVT151012 formed at \response{$Z = 10\% \solar$}. The grey diagonal dashed line shows $M_1^\mathrm{BH} = M_2^\mathrm{BH}$. The constraints we use to determine if a merging binary black hole is similar to one of the observed \ac{GW} events are shown in grey and described in Results.
	
(c) \ac{ZAMS} masses $M_1^\mathrm{ZAMS}$ and $M_2^\mathrm{ZAMS}$ for GW150914, GW151226 and LVT151012 progenitors at \response{the lower metallicity $Z=\fivepercentsolar=0.001$}. The progenitor masses required to produce GW151226 and LVT151012 decrease, and we are able to produce GW150914. 
	
(d) Final black hole masses $M_1^\mathrm{BH}$ and $M_2^\mathrm{BH}$ for GW150914, GW151226 and LVT151012 \acp{BBH} formed from \response{5\%-solar metallicity progenitors.}
	
	The panels of this figure are formatted to be comparable to Figure 4 in \citet{BBH:O1}.}
	\label{fig:event-progenitor-masses}
\end{figure*}

We simulate events at \response{10\%-solar ($Z=0.002$) and 5\%-solar ($Z=0.001$) metallicity} using the \texttt{Fiducial} model assumptions (see Methods).  We select binaries which fall within the 90\% credible interval on total (chirp) \ac{BBH} mass and with $q$ above the 90\% credible interval lower bound for GW150914 (GW151226 and LVT151012).  In all cases, we select only \acp{BBH} that merge within the Hubble time. Systems satisfying these conditions are shown in Figure~\ref{fig:event-progenitor-masses}. The upper panel shows \acp{BBH} formed at \response{10\%-solar metallicity} whilst the lower panel shows those formed at \response{5\%-solar metallicity}. The black hole mass of the initially more massive star is labeled as $M_1^\mathrm{BH}$ and that of the initially less massive star  as $M_2^\mathrm{BH}$. 

In the left hand column of Figure~\ref{fig:event-progenitor-masses} we show the \ac{ZAMS} masses of possible progenitors of these events. Progenitors of the events are separated in \ac{ZAMS} masses apart from rare systems that start on very wide orbits, avoiding mass transfer altogether, but are brought to merger by fortuitous supernova kicks.  These systems do not lose mass through non-conservative mass transfer, and can therefore form more massive binaries from lower mass progenitors -- the LVT151012 outlier progenitor in the lower left corner of the bottom left panel of Figure~\ref{fig:event-progenitor-masses} was formed this way.

\response{Massive stars have high mass loss rates; e.g., at solar metallicity, massive stars could lose tens of solar masses through winds even before interacting with their companion.}   We find, in agreement with \citet{GW150914:astro} and \citet{Belczynski:2016}, that it is not possible to form GW150914 or LVT151012 through classical isolated binary evolution at solar metallicity.  \response{GW151226 lies at the high-mass boundary of \acp{BBH} that can be formed at solar metallicity.}

\response{GW151226 is consistent with being formed through classical isolated binary evolution at 10\%-solar metallicity from a binary with total mass $65 \lesssim M / M_\odot \lesssim 100$ (see upper left panel of  Figure~\ref{fig:event-progenitor-masses}).  LVT151012 is also consistent with being formed at 10\%-solar metallicity from binaries with initial total masses $95 \lesssim M / M_\odot \lesssim 125$.  Typical progenitors have a mass ratio close to unity (median $q=0.75$), with an initial orbital period of $\sim 500$ days.}

GW150914 could have formed through isolated binary evolution at metallicities \response{$Z \lesssim \fivepercentsolar$} from binaries with initial total mass \response{$\gtrsim 160 \Msol$} (see lower left panel of Figure~\ref{fig:event-progenitor-masses}).  While this mass range is similar to that found by others who investigated the formation of GW150914 through isolated binary evolution at low metallicities \citep{Belczynski:2016, Lipunov:2016,EldridgeStanway:2016}, we note that, unlike \citet{EldridgeStanway:2016}, we do not require fortuitous supernova kicks resulting in high eccentricity to form this binary at \response{$Z=\fivepercentsolar$}.  We identify the same main evolutionary channel (see Figure~\ref{fig:typical-progenitors}) as \citet{Belczynski:2016}.  \response{We find that GW151226 and LVT151012 are also consistent with forming through this channel at lower metallicity, from initially lower mass binaries. For example, the total progenitor binary mass range for forming GW151226 reduces from $65 \lesssim M / M_\odot \lesssim 100$ at 10\% solar metallicity to $60 \lesssim M / M_\odot \lesssim 90$ at 5\%-solar metallicity, demonstrating a degeneracy in the \ac{ZAMS} masses and metallicity inferred in our model due to the dependence of mass loss rates on metallicity.}

We find that the chirp masses of GW151226 and LVT151012 lie near the peak of the mass distribution of \ac{BBH} mergers formed at 10\%-solar metallicity which are observable by aLIGO. There remains significant support for both systems at 5\%-solar metallicity. GW150914 cannot be formed at 10\%-solar metallicity in our model, and remains in the tail of the total mass distribution at 5\%-solar, which is the highest metallicity at which we form significant numbers of all three event types in the \texttt{Fiducial} model. Events like GW150914 are much more common at 1\%-solar metallicity.

\response{At $Z=\fivepercentsolar$, the more massive black hole is formed from the initially more massive star in $\sim 90\%$ of systems.}

\response{Interestingly, low metallicities can produce significantly unequal mass ratios.  For example, the median mass ratio of merging \acp{BBH} is $\sim 0.5$ at 10\% solar metallicity.}  The high  fraction of merging \acp{BBH} with low mass ratios at low metallicities is a general trend; this agrees with Figure 9 of \citet{Dominik:2012}, who do not, however, discuss this effect. A \ac{GW} detection of a heavy \ac{BBH} with an accurately measured low mass ratio could indicate formation in a lower metallicity environment, and not necessarily dynamical formation as suggested in \citet{BBH:O1}.

The significant fraction of low mass-ratio mergers at low metallicity arises due to a combination of effects. The maximum \ac{BH} mass for single stars is a function of metallicity (e.g., Figure 6 of  \citet{Spera:2015}), with more massive \acp{BH} formed at lower metallicities due to reduced mass loss.  Therefore, for a given observed chirp mass, more unequal \acp{BH} can be formed at low metallicity.  A second effect comes from the difference in the onset of the first episode of mass transfer, which is key for determining the mass of the remnant.  The dependence of stellar radius on metallicity \citep{Pols:1998} means that stars with lower metallicity experience their first episode of mass transfer in a more evolved phase of their evolution for a given initial orbital separation \citep{deMink:2007}.  They thus lose less mass when the hydrogen envelope is stripped, again allowing for more unequal remnants. 

\subsection*{Typical evolutionary pathway of GW151226}
\label{subsec:pathway}

\begin{figure*}
\centering
\includegraphics[clip, width=\textwidth, trim=4cm 4cm 4cm 1cm]{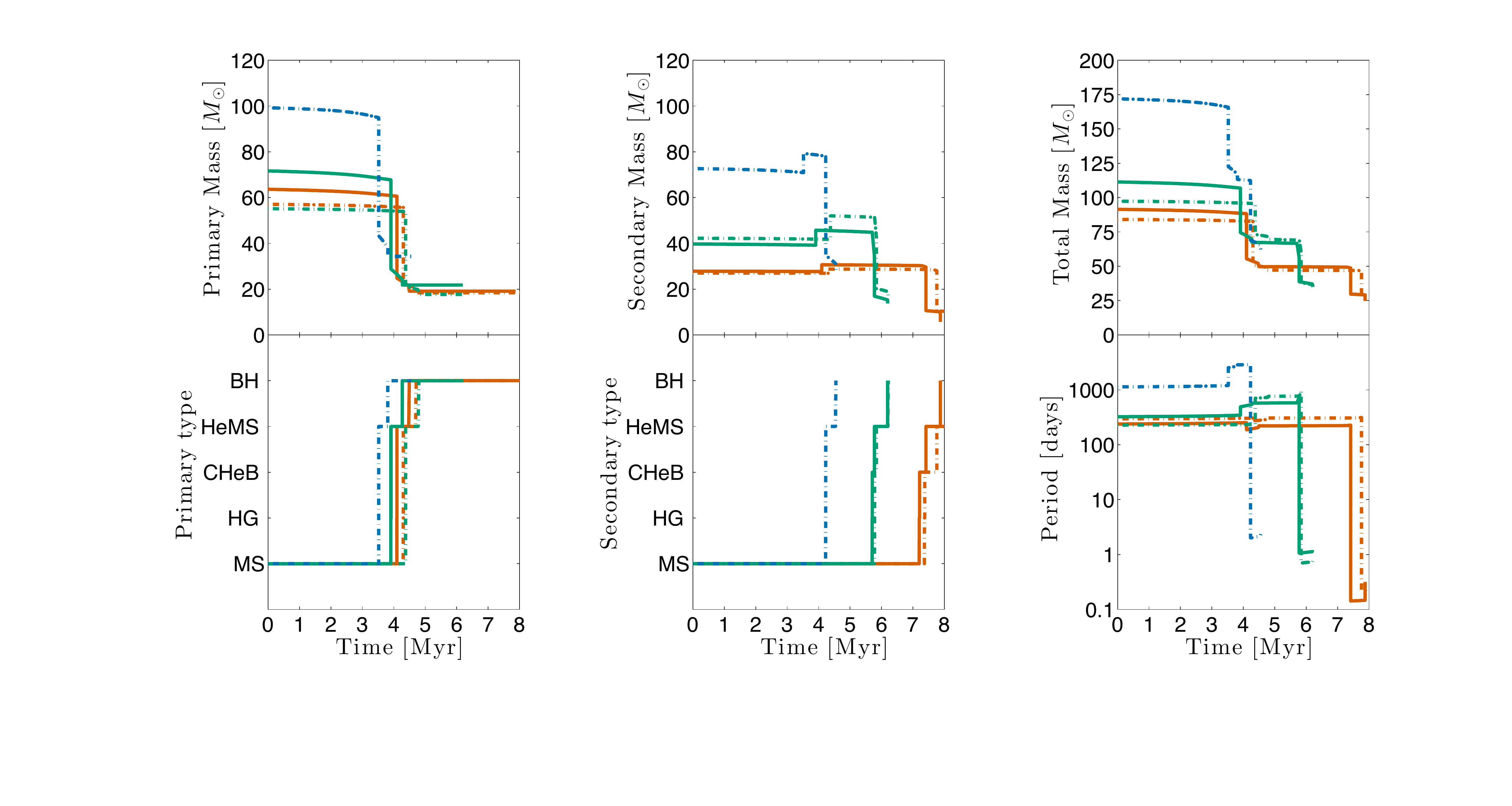}
\caption[]{\textbf{Typical evolution of BBH progenitors.} 

Evolution in time of representative GW150914 (blue), GW151226 (orange) and LVT151012 (green) progenitors at \response{10\%-solar ($Z=0.002$, solid lines) and 5\%-solar metallicity ($Z=0.001$, dashed lines). }
(a) The mass of the initially more massive star. The stars lose mass through stellar winds, mass transfer and supernovae.
(b) The mass of the secondary star. The stars may accrete mass during mass transfer episodes.
(c) The evolution of the total mass of the binary.
(d) The evolutionary stage (stellar type) of the initially more massive star as given by \citet{Hurley:2000} (see Results for definitions). 
(e) The evolutionary stage (stellar type) of the secondary star. 
(f) The orbital period of the binary in days.}
\label{fig:typical-progenitors}
\end{figure*}

\begin{figure*}
\centering
\includegraphics[clip, width=\textwidth, trim=0cm 10cm 0cm 0cm]{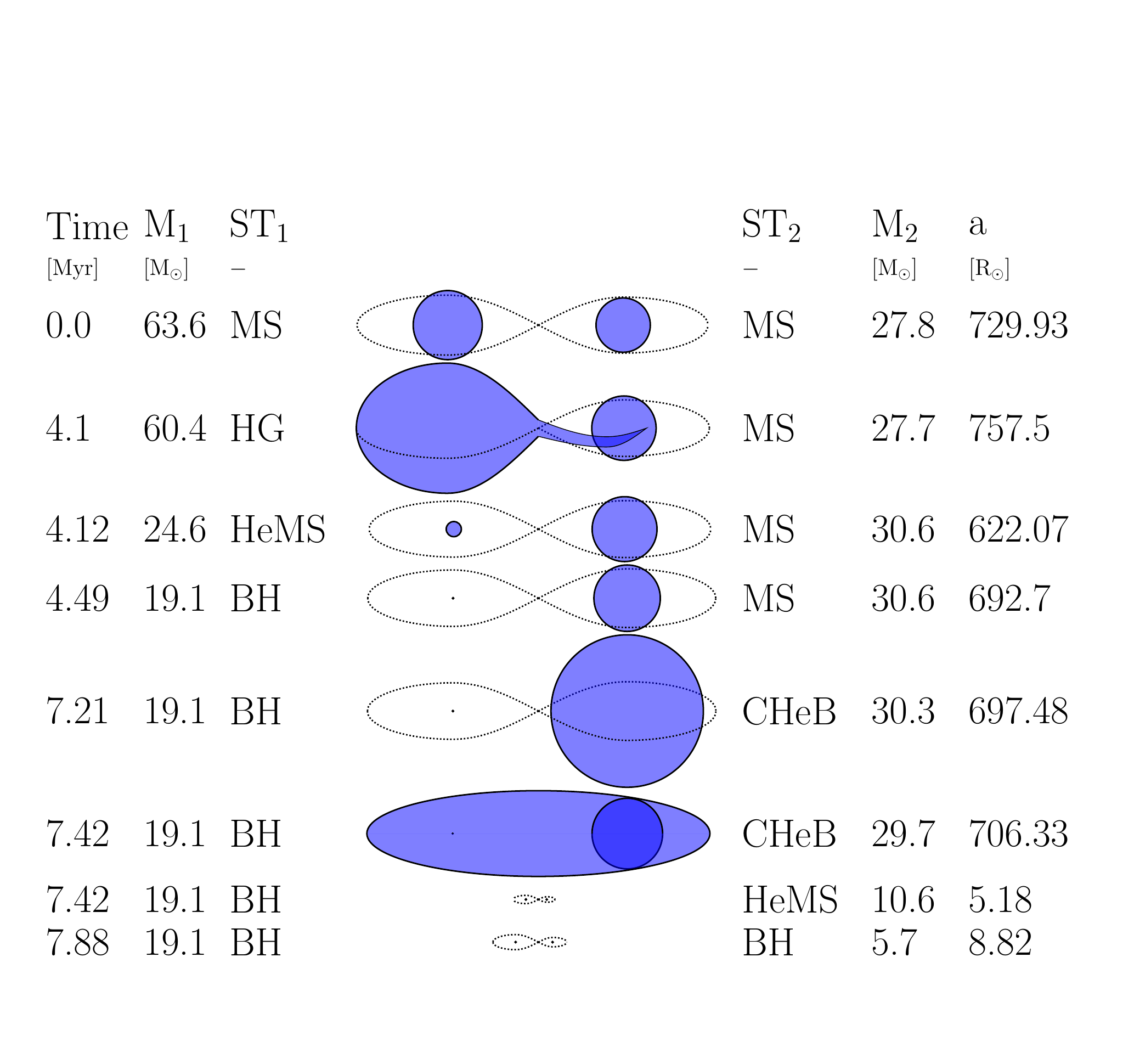}
\caption[]{\textbf{Formation of GW151226}

\response{Typical formation of GW151226 at 10\%-solar metallicity in our model, as described in the Results. The columns show the time, the masses and stellar types of the primary and secondary, $\mathrm{M}_1$, $\mathrm{ST}_1$ and $\mathrm{M}_2$, $\mathrm{ST}_2$ respectively, and the semi-major axis $a$. Some intermediate stages of the evolution are omitted for clarity.}}
\label{fig:vdH}
\end{figure*}

\response{In Figure~\ref{fig:typical-progenitors} we show the evolution in time of the masses, stellar types and orbital period of typical progenitors of all three observed \ac{GW} events. Progenitors of all
three systems follow the same typical channel.  Here we describe the evolution of a
typical 10\%-solar metallicity progenitor of GW151226 (solid orange line in Figure~\ref{fig:typical-progenitors}); it is shown graphically in Figure~\ref{fig:vdH}.}

\response{The binary initially has two high-mass main-sequence (MS) O stars, a primary
of $\sim 64 M_{\odot}$ and a $\sim 28 M_{\odot}$ companion with an initial orbital period of $\sim 300$ days.
The primary expands at the end of its main sequence evolution, fills its Roche
lobe and initiates mass transfer as a $\sim 60 M_{\odot}$ \ac{HG} or \ac{CHeB} star (case B or C mass transfer), donating its $\sim 36 M_{\odot}$
hydrogen-rich envelope to the secondary, which accretes only $\sim 3 M_{\odot}$ of
it. This leaves the primary as a stripped naked \ac{HeMS} of $\sim 25 M_{\odot}$.
After evolving and losing a few solar masses through stellar winds, the primary
collapses to a \ac{BH} of $\sim 19 M_{\odot}$ through almost complete fallback.}

\response{The secondary continues evolving and initiates mass transfer as a \ac{CHeB} star
of $\sim 30 M_{\odot}$. This mass transfer is dynamically unstable and leads to the formation
and subsequent ejection of a \ac{CE}. The \ac{CE} ejection draws energy from the orbit and
results in significant orbital hardening: the orbital period is reduced by $\sim 3$ orders
of magnitude as can be seen in the lower right panel of Figure~\ref{fig:typical-progenitors}. The secondary, which becomes a \ac{HeMS} star of $\sim 11 M_{\odot}$ after the ejection of the envelope, eventually collapses to a $\sim 6 M_{\odot}$ \ac{BH}. Finally, the binary merges through \ac{GW}  emission in $\sim 100$ Myrs.}

\response{A few percent of our \ac{BBH} progenitors form through a variant of this channel involving a double \ac{CE}.  This variant involves two nearly equal mass \ac{ZAMS} stars which first interact during the \ac{CHeB} phase of their evolution, initiating a double \ac{CE} which brings the cores close together.  This is followed by both stars collapsing into \acp{BH} and merging through \ac{GW} emission.}

\section*{Discussion}
\label{sec:discussion}

We have explored whether all of the \ac{GW} events observed to date could have been formed through classical isolated binary evolution via a \ac{CE} phase.  All three observed systems can be explained through this channel under our \texttt{Fiducial} model assumptions.  Forming all observed \ac{GW} events through a single formation channel avoids the need to fine tune the merger rates from the very different evolutionary channels discussed in the Introduction to be comparable. Other proposed formation scenarios struggle to produce at least one of the observed \acp{BBH}.  For example, both chemically homogeneous evolution \citep{MandeldeMink:2016,Marchant:2016,deMinkMandel:2016} and dynamical formation in old, low-metallicity globular clusters in the model of \citet{Rodriguez:2016} (see their figure 2) have little or no support for relatively low-mass \acp{BBH} such as 
GW151226, which has a total mass $M = 21.8 \pm_{1.7}^{5.9} M_\odot$ \citep{BBH:O1}. The ability of a single channel to explain all observed events will be tested with future \ac{GW} observations \citep{BBH:O1,scenarios}.

\response{We form $\sim 2 \times 10^{4}$ \acp{BBH} that merge in a Hubble time per $1 \times 10^{9}$ solar masses of star formation at 10\%-solar metallicity in our \texttt{Fiducial} model, using the \citet{Kroupa:2001} \ac{IMF}, a uniform mass ratio distribution and assuming that all stars are in binaries. This increases to $\sim 3 \times 10^{4}$ \acp{BBH} per $1 \times 10^{9}$ solar masses of star formation at $Z=\fivepercentsolar$. Rescaling by the total star formation rate\citep{MadauDickinson:2014} at redshift $z=0$ , this would correspond to a \ac{BBH} formation rate of $\sim 300$ Gpc$^{-3}$ yr$^{-1}$ assuming all star formation happens at 10\%-solar metallicity. This can be compared to the empirical LIGO \ac{BBH} merger rate estimate\citep{BBH:O1} of 9 -- 240 Gpc$^{-3}$ yr$^{-1}$. However, this comparison should be made with caution, because even local mergers can arise from binaries formed at a broad range of redshifts and metallicities.  An accurate calculation of the merger rate requires the convolution of the metallicity-specific redshift-dependent star formation rate with the time delay distribution, integrated over a range of metallicities\citep{Dominik:2014}.  }

There are many uncertainties in the assumptions we make (see Methods for details of our default assumptions). The evolution of massive progenitor binaries is poorly constrained by observations, although there has been recent progress, such as with the VLT-FLAMES Tarantula Survey (VFTS) in the 30 Doradus region of the Large Magellanic Cloud\citep{Sana:2013VFTS}. 

In rapid population synthesis codes like COMPAS, these uncertainties are treated by parametrising complex physical processes into simple one or two parameter models, such as treating the \ac{CE} with the $\alpha$ prescription \citep{Webbink:1984}, or scaling \ac{LBV} mass loss rates with $f_\mathrm{LBV}$. The multidimensional space of model parameters, including $\alpha$  and $f_\mathrm{LBV}$, must then be explored in order to properly examine the model uncertainties. 

We leave a full exploration of this parameter space for future studies with COMPAS; here we follow the common approach\citep{Dominik:2012,Mennekens:2014,deMinkBelczynski:2015} of varying individual parameters independently and assessing their impact relative to the \texttt{Fiducial} model.  

In the \texttt{Fiducial} model, we used the `delayed' supernova model of \citet{Fryer:2012}. We have also checked that using the `rapid' model of \citet{Fryer:2012} does not significantly alter the typical evolutionary pathways for forming heavy \acp{BBH} discussed here, since both models predict high-mass \ac{BH} formation through almost complete fallback.

\citet{Mennekens:2014} use a \ac{LBV} mass loss rate of $10^{-3}  M_\odot$ $ \mathrm{yr}^{-1}$. They find that such strong mass loss can shut off the typical channel for \ac{BBH} formation. In COMPAS, we follow  \texttt{SSE} \citep{Hurley:2000} for identifying LBVs as massive stars with $L/L_{\odot} > 6\times10^5$ and $(R/R_{\odot})(L/L_{\odot})^{1/2} > 10^{5}$. We find that increasing the mass loss rate of LBVs from $1.5 \times 10^{-4} M_\odot$ to $10^{-3} M_\odot$ $\mathrm{yr}^{-1}$ does not significantly change the total \ac{BBH} merger rate; nevertheless, the number of \ac{BBH} mergers similar to LVT151012 was reduced by a factor of $\sim 10$ for progenitors at 5\%-solar metallicity.

In the \texttt{Fiducial} model we only permit evolved \ac{CHeB} stars with a well defined core-envelope separation to survive \ac{CE} events (see Methods). This model therefore corresponds to the pessimistic model of \citet{Dominik:2012}, which is also the standard model (M1) of \citet{Belczynski:2016}. We also consider an alternate model where we allow \ac{HG} donors to initiate and survive \ac{CE} events, as in the optimistic model of \citet{Dominik:2012}. \response{We find that the optimistic \ac{CE} treatment predicts total \acp{BBH} merger rates which are $\sim 3$ times higher than the \texttt{Fiducial} model at $Z=\tenpercentsolar$, and $\sim 2$ times higher at $Z=\fivepercentsolar$. This optimistic variation also raises the total merging \ac{BBH} mass that can be formed at a given metallicity; e.g., at  $Z=\tenpercentsolar$, the maximum total \ac{BBH} mass rises from $\sim 50 M_\odot$ for the pessimistic model to $\sim 60 M_\odot$ for the optimistic model, as also noted by \citet{Dominik:2012}.}  The spread between these optimistic and pessimistic models also reflects the uncertainty in the radial evolution of very massive stars; the results of the pessimistic model could move toward those of the optimistic model if the radial expansion for the most massive stars predominantly happens during the \ac{CHeB} phase rather than during the \ac{HG} phase.

For a very small number of our simulated systems, immediately after the \ac{CE} is ejected the binary is comprised of a \ac{BH} and a \ac{HeMS} secondary that is already overfilling its Roche lobe.  In the \texttt{Fiducial} model we treat these systems as an unsuccessful \ac{CE} event, leading to mergers. Similar studies \citep{deMink:2007SMC,Podsiadlowski:2010} have allowed only those systems which overfill the Roche lobe by no more than $10\%$ at the end of the \ac{CE} phase to survive. We also consider the extreme alternative of allowing all such systems to survive. The \ac{HeMS} stars lose a significant fraction of their mass through rapid but stable mass transfer onto the \ac{BH} companion.  Most of this mass is removed from the binary as the \ac{BH} companion can only accrete at the Eddington limit, and the \ac{HeMS} star leaves behind a relatively low mass \ac{BH}.  We verify that this has no impact on our conclusions.

\response{We test the impact of the assumed \ac{CE} ejection efficiency by changing the value of $\alpha\lambda$ from the fiducial $0.1$ to $0.01$. At 10\%-solar metallicity we find the total \ac{BBH} merger rate drops by a factor of $\sim 2$. \citet{Dominik:2012} performed the same study, setting $\alpha\lambda = 0.1$ (model V2) and $\alpha\lambda = 0.01$ (model V1) and report the same decrease (see tables 1,2 and 3 in \citet{Dominik:2012}). At 5\%-solar metallicity, the total \ac{BBH} merger rate drops by a factor of $\sim 4$, with the specific merger rates of binaries like GW151226, LVT151012, and GW150914 dropping by factor of $\sim 25$, $\sim 4$, and $\sim 50$, respectively. The maximum \ac{BBH} mass produced at 10\%-solar metallicity increases from $\sim 50 M_\odot$ in the \texttt{Fiducial} model to $\sim 60 M_\odot$ under this variation. At 5\%-solar metallicity we find that the maximum total \ac{BBH} mass decreases from $\sim 75 M_\odot$ to $\sim 65 M_\odot$. }

In conclusion, we have shown that GW150914, GW151226 and LVT151012 are all consistent with formation through the same classical isolated binary evolution channel via mass transfer and a common envelope.  \ac{GW} observations can place constraints on the uncertain astrophysics of binary evolution \citep{BulikBelczynski:2003,OShaughnessy:2013,Stevenson:2015,Mandel:2015,Mandel:2016prl}.  Although the focus of this paper has been on the constraints placed by the observed \ac{BBH} masses, other observational signatures, including merger rates (and their variation with redshift) \citep{MandelOShaughnessy:2010}, \ac{BH} spin magnitude and spin-orbit misalignment measurements \citep{Vitale:2015tea,Kushnir:2016,Rodriguez:2016vmx}, and possibly a \ac{GW} stochastic background observation \citep{GW150914:stoch,Callister:2016}, can all contribute additional information.  COMPAS will provide a platform for exploring the full evolutionary model parameter space with future \ac{GW} and electromagnetic observations.

\section*{Acknowledgments}

We would like to thank Chris Belczynski, Christopher Berry, Natasha Ivanova, Stephen Justham, Vicky Kalogera, Gijs Nelemans, Philipp Podsiadlowski, David Stops and Alberto Vecchio for useful discussions and suggestions.  IM acknowledges support from STFC grant RRCM19068.GLGL; his work was performed in part at the Aspen Center for Physics, which is supported by National Science Foundation grant PHY-1066293. AVG acknowledges support from CONACYT. SS and IM are grateful to NOVA for partially funding their visit to Amsterdam to collaborate with SdM.  SdM acknowledges support by a Marie Sklodowska-Curie Action (H2020 MSCA-IF-2014, project id 661502) and National Science Foundation under Grant No. NSF PHY11-25915. 

\section*{Author contributions}

All authors contributed to the analysis and writing of the paper.

\section*{Data Availability}

We make the results of our simulations available at \url{http://www.sr.bham.ac.uk/compas/}.

\section*{Competing Financial Interests}

The authors declare no competing financial interests.

\section*{Methods}
\subsection*{COMPAS population synthesis code}
\label{sec:COMPAS}

\ac{COMPAS} includes a rapid Monte-Carlo binary population synthesis code to simulate the evolution of massive stellar binaries, the possible progenitors of merging compact binaries containing \acp{NS} and \acp{BH} which are potential \ac{GW} sources.  Our approach to population synthesis is broadly similar to \texttt{BSE} \citep{Hurley:2002} and the codes derived from it, such as \verb|binary_c| \citep{Izzard:2004,Izzard:2006,Izzard:2009,deMink:2013} and \texttt{StarTrack} \citep{Belczynski:2002,Belczynski:2008}.  

\ac{COMPAS} was developed to explore the many poorly constrained stages of binary evolution, such as mass transfer, \ac{CE} evolution and natal supernova kicks imparted to \acp{NS} and \acp{BH} \citep{PostnovYungelson:2014}.  Here we provide a brief overview of our default assumptions.

For our \texttt{Fiducial} model, we simulate likely \acp{BBH} progenitor binaries with the primary mass $m_1$ drawn from the Kroupa \ac{IMF} \citep{Kroupa:2001} up to  $m_1 \leq 100  M_\odot$  where the \ac{IMF} has a power-law index of $-2.3$.  The mass of the secondary is then determined by the initial mass ratio $q \equiv m_2 / m_1$, which we draw from a flat distribution between 0 and 1 \citep{Sana:2012}.  

The semimajor axis $a$ is chosen from a flat-in-the-log distribution \citep{Opik:1924,Abt:1983} and restricted between $0.1 < a / \mathrm{AU} < 1000$; the period distribution is therefore set by the convolved semimajor axis and mass distributions.  The boundaries on the component masses and separations are chosen to safely encompass all individual solutions yielding \acp{BBH} of interest, and so impact normalisation only.   Binaries are assumed to have an initial eccentricity of zero; the initial semimajor axis distribution serves as a proxy for the periapsis distribution, which is the relevant parameter affecting binary evolution \citep{deMinkBelczynski:2015}. Stellar rotation and tides are not included in the \texttt{Fiducial} model.

We use the analytical fits of \citet{Hurley:2000} to the models of \citet{Pols:1998} for single stellar evolution. We note that the original grid of single star models extends only to 50 solar masses. We extrapolate above this limit, as described in \citet{Hurley:2000}.

We include mass loss due to stellar winds for hot O stars following the Vink model \citep{Vink:2001, Belczynski:2010}, with a \ac{LBV} mass loss rate of $f_\mathrm{LBV} \times 10^{-4} \Msol$ $ \textrm{yr}^{-1}$, independent of metallicity. In the \texttt{Fiducial} model $f_\mathrm{LBV} = 1.5$ \citep{Belczynski:2010}. For \ac{WR} stars we use the formalism of \citet{Hamann:1998}, modified as in \citet{Belczynski:2010} to be metallicity dependent ($\propto Z^{0.85}$) based on \citet{Vink:2005}. We assume that all stellar winds are isotropic and remove the specific angular momentum of the mass losing object. We do not account for wind accretion by a companion.

Mass transfer occurs when the donor star fills its Roche lobe, whose radius is calculated according to \citet{Eggleton:1983}. Although all of our binaries are initially circular, supernovae can lead to some eccentric systems.  We use the periastron to check whether a star would fill its Roche lobe, whose radius is computed for a circular orbit with the periastron separation. We assume that mass transfer circularises the orbit.

In the absence of accurate stellar models spanning the full parameter space of interest, we use a simplified treatment of mass transfer.  We assume that mass transfer from main-sequence, core-hydrogen-burning donors (case A) is dynamically stable for mass ratios $q \geq 0.65$.  We follow \citet{deMink:2013} and \citet{Claeys:2014} in assuming that case A systems with $q < 0.65$ will result in mergers as the accretor expands and brings the binary into contact \citep{deMink:2007SMC}.  Stable case A mass transfer is solved using an adaptive algorithm \citep{Schneider:2015} which requires the radius of the donor to stay within its Roche lobe during the whole episode; when this is impossible, we assume that any donor mass outside the Roche lobe is transferred on a thermal timescale until the donor is again contained within its Roche lobe.
In our Fiducial model we first test whether mass transfer is stable; if it is, we treat stable mass transfer from all evolved stars (case B or case C) equally, without distinguishing between donors with radiative and convective envelopes: we remove the entire envelope of the donor on its thermal timescale \citep{KippenhahnWeigert:1967}.
We follow \citet{Tout:1997,Belczynski:2008} in our model for the rejuvenation of mass accreting stars.

The efficiency of mass transfer (i.e., how conservative it is) is set by the rate at which the accretor can accept material from the donor. For \ac{NS} and \ac{BH} accretors, the maximum rate of accretion is defined by the Eddington limit.  We assume that a star can accrete at a rate $C M_\textrm{acc}/\tau_\mathrm{th}$, with the Kelvin-Helmholtz thermal timescale $\tau_\mathrm{th} = GMM_\textrm{env}/RL$, where $G$ is the gravitational constant, $M$ is the total mass of the star, $M_\mathrm{env}$ is the mass of the envelope, $R$ is the radius of the star and $L$ is its luminosity. 
The constant $C$ is a free parameter in our model; we use $C=10$ for all accretion episodes in the \texttt{Fiducial} model\cite{Hurley:2002}.
The material that fails to be accreted is removed from the system with the specific angular momentum of the accretor via isotropic re-emission.

We determine the onset of dynamically unstable mass transfer by comparing the response of the radius of the donor star to a small amount of mass loss against the response of the orbit to a small amount of mass transfer \citep{Soberman:1997}. We use fits to condensed polytrope models  \citep{Hjellming:1987,Soberman:1997} to calculate the radius response of a giant to mass loss on a dynamical timescale. Dynamically unstable mass transfer leads to a \ac{CE}. If the donor star is on the \ac{HG}, we follow \citet{Belczynski:2007,Belczynski:2016} in assuming such systems cannot survive a \ac{CE}. In fact, such systems may never enter CE at all. \citet{Pavlovskii:2016} have shown that in many cases mass transfer from \ac{HG} donors will be stable and not lead to a \ac{CE}.

All of our successful \ac{CE} events therefore involve a donor star which has reached \ac{CHeB}. For \ac{CE} events, the $\lambda$ parameter, which characterizes the binding energy of the envelope \citep{Webbink:1984}, is set to $\lambda = 0.1$ \citep{DewiTauris:2000,Loveridge:2011,Dominik:2012,Belczynski:2016} while the $\alpha$ parameter, which characterizes the efficiency of converting orbital energy into \ac{CE} ejection, is set to $\alpha = 1$.  If one of the stars in the post-\ac{CE} binary is filling its Roche lobe immediately after CE ejection, we assume that there is insufficient orbital energy available to eject the envelope and the binary evolution is terminated in a merger.  We assume that \ac{CE} events with successful envelope ejections circularise orbits (see section 10.3.1 of \citet{Ivanova:2013}.)

The relationship between the pre-supernova core mass and the compact remnant mass follows the `delayed' model of \citet{Fryer:2012}. Supernova kicks are assumed to be isotropic and their magnitude is drawn from a Maxwellian distribution with a 1D velocity dispersion $\sigma=250$ km s$^{-1}$ \citep{Hobbs:2005}, reduced by a factor of $(1-f)$, where $f$ is the fallback fraction, calculated according to \citet{Fryer:2012}.  As in \citet{Belczynski:2016}, we find that most of our heavy black holes form through complete fallback without a supernova or associated kick. 

\bibliographystyle{customnature}
\bibliography{Mandel_edit}

\label{lastpage}

\end{document}